\documentclass[aps,pra,twocolumn,superscriptaddress]{revtex4-1}
\usepackage{epsfig,pstricks,graphicx}
\usepackage{amsmath,amssymb}
\usepackage[mathscr]{eucal}
\usepackage{color}
\usepackage{comment}
\usepackage[normalem]{ulem}

\newcommand{\tcb}[1]{\textcolor{blue}{#1}}

\def\CC{{\rm\kern.24em \vrule width.04em height1.46ex depth-.07ex
\kern-.30em C}}
\def\RR{{\rm
         \vrule width.04em height1.58ex depth-.0ex
         \kern-.04em R}}
\def\P{{\rm I\kern-.25em P}}
\def\id{{\rm 1\kern-.22em l}}

\def\argmax{\qopname\relax n{argmax}}
\def\argmin{\qopname\relax n{argmin}}

\newcommand{\bra}[1]{\left\langle #1 \right |}
\newcommand{\ket}[1]{\left | #1 \right\rangle}

\newcommand{\oost}{\frac{1}{\sqrt{2}}}
\newcommand{\rme}{\ensuremath{\mathrm{e}}}
\newcommand{\rmi}{\ensuremath{\mathrm{i}}}
\newcommand{\rmd}{\ensuremath{\mathrm{d}}}

\newcommand{\tr}{\operatorname{tr}}

\newcommand{\rhoS}{\rho^{\mathrm{S}}}

\begin{document}
\title{ Practical method to obtain a lower bound to the three-tangle
      }
\author{Christopher Eltschka}
\affiliation{Institut f\"ur Theoretische Physik, 
         Universit\"at Regensburg, D-93040 Regensburg, Germany}
\author{Jens Siewert}
\affiliation{Departamento de Qu\'{\i}mica F\'{\i}sica, Universidad del Pa\'{\i}s Vasco UPV/EHU,
             E-48080 Bilbao, Spain}
\affiliation{IKERBASQUE, Basque Foundation for Science, E-48011 Bilbao, Spain}

\begin{abstract}
   The quantitative assessment of the entanglement in multipartite quantum 
   states
   is, apart from its fundamental importance, a practical problem.
   Recently there has been significant progress in developing new methods
   to determine certain entanglement measures. In particular, there is 
   a method---in principle, analytical---to compute a certified lower bound
   for the three-tangle.
   The purpose of this work is to provide a manual for the 
   implementation of this approach
   and to explicitly discuss several analytically solvable cases 
   in order to gauge 
   the numerical tools. Moreover, we derive a simple analytical bound
   for the mixed state three-tangle.
\end{abstract}
\maketitle

\section{Introduction}

The quantitative theory of multipartite 
entanglement~\cite{Plenio2007,Amico2008,Horodecki2009} 
is far from being a mature subject. The three-qubit problem, i.e., 
the quantification of Greenberger-Horne-Zeilinger (GHZ) entanglement in 
an arbitrary three-qubit state,
may serve as an acid test for progress in this field. 
Recent years have brought substantial advancements
regarding methods to estimate entanglement 
measures~\cite{Horodecki1999,Audenaert2006,Guehne2007,Eisert2007,OSU2008,Loss2008,OH2010,Cao2010,Jungnitsch2011,ChenHuber2011,Kampermann2012,Chen2012,
Zhu2012,Love2013} 
on the one hand, and in terms of exact solutions of very specific 
problems~\cite{LOSU2006,EOSU2008,Jung2009-1,Jung2009-2,Gour2010,Fei2011,Viehmann2012,SE2012} 
on the other hand.
Although a mathematical insider of such methods would be able to produce 
lower bounds on the GHZ-type entanglement in some generic cases the problem for the 
practitioner has remained open: Given an arbitrary three-qubit mixed state, e.g., the 
density matrix of an experimentally prepared state 
(such as in recent 
experiments~\cite{Martinis2010,Schoelkopf2010,Fedorov2012,Blatt2012}),
determine the quality of the GHZ-type entanglement.

The objective of this article is to fill this gap. In a recent publication~\cite{ES2012-ScR} 
we have proposed a method to quantify GHZ-type entanglement in arbitrary $N$-qubit  states.
The three-qubit case was spelled out in some detail, however, not at the level 
at which
one could directly apply it without some in-depth study of the background.
In contrast, the present work focuses on the practical application of these
findings. We explain the ingredients of the method and the steps one has to 
go through in order to implement the approach. Moreover, we provide an approximate
analytical formula as well as a basic error estimate.

Why is it so important to obtain a {\em lower} bound? As the three-tangle for
mixed states---like many other entanglement measures---is defined as a convex
roof (see below) it is a {\em minimum} taken over all possible decompositions
of a given state.
Hence, any decomposition gives an {\em upper} bound, however, it is difficult
to say by how much this bound overestimates the true three-tangle. In particular,
knowing only upper bounds it cannot be excluded that the three-tangle 
vanishes. Therefore it is essential to obtain a lower bound.  Once a 
non-trivial lower bound
is known it becomes meaningful to look for good upper bounds as well.

The article is organized as follows: In Section II, we start by outlining the 
method and the most important concepts it requires. These concepts are then 
explained in separate parts: GHZ-symmetric states and the quantification of
their three-tangle (Section III), symmetrization of 
an arbitrary state (Section IV) and
analytical estimation of  its three-tangle,
 and entanglement optimization (Section V). In Section VI
we highlight some experimental aspects of our method, and in 
Section VII  we present an error estimate. Finally, in Section VIII
we discuss the performance of our approach by applying it to several 
exactly solvable cases of three-qubit states/families.

\section{The method and its ingredients}

In this section we introduce all the concepts and ideas that are
required to solve our task -- to estimate the amount of GHZ-type
entanglement in three-qubit states. While we just mention
them here they are be considered and explained in detail
in the following sections.

We exclusively consider three-qubit
mixed states $\rho\in \mathcal{B}(\mathcal{H})$, that is, 
bounded positive definite Hermitian
operators that act on the Hilbert space of three qubits 
$\mathcal{H}=\CC^2\otimes\CC^2\otimes\CC^2$.
Quantification of the entanglement in $\rho$ makes sense only
if the states are normalized: $\tr\rho=1$. Sometimes, however,
we encounter states that are not normalized.

Suppose we are given a generic three-qubit state $\rho$
and would like to determine how much GHZ-type entanglement
it contains. The appropriate entanglement measure for this
purpose is the
three-tangle~\cite{CKW2000,Verstraete2003,Gour2010,Viehmann2012},
denoted by $\tau_3$. It is the square root of the {\em residual tangle} 
originally introduced by Coffman {\em et al.}~\cite{CKW2000}.

A lower bound to the exact $\tau_3(\rho)$ can be found by
proceeding according to the following recipe:
\begin{enumerate}
  \item[{\em (1)}] 
        Find the so-called {\em normal form}
        of $\rho$.  We denote the normal form 
        $\rho^{\mathrm{NF}}$. As this operation in general is non-unitary we have
        $\tr\rho^{\mathrm{NF}}\leqq 1$.\\
        If $\tr\rho^{\mathrm{NF}}=0$, the procedure terminates here and $\tau_3(\rho)=0$.
  \item [{\em (2)}] 
        Optimize $\rho^{\mathrm{NF}}/\tr\rho^{\mathrm{NF}}$ by applying local unitary 
        operations to the qubits according to an appropriate criterion (see below). 
        The result is $\tilde{\rho}^{\mathrm{NF}}$.
  \item [{\em (3)}] 
        Project $\tilde{\rho}^{\mathrm{NF}}$ onto the GHZ-symmetric states and
        read off the value for $\tau_3$. The result 
        $\tau_3\left(\mathbf{P}(\tilde{\rho}^{\mathrm{NF}})\right)$ 
        leads to the desired lower bound:
\[
        \tau_3\left(\mathbf{P}(\tilde{\rho}^{\mathrm{NF}})\right) \tr \rho^{\mathrm{NF}}\ \leqq \ \tau_3(\rho) \ \ \ .
\]
\end{enumerate}
The idea behind this sequence is simple. In step~{\em (3)}, the state is 
projected onto a family of symmetric states for which the exact three-tangle is
known. In this projection, generally
entanglement is lost, but never gained. In order to minimize
the entanglement loss one has to optimize the state -- which is the purpose
of steps {\em (1)} and {\em (2)}.
\section{Three-tangle of GHZ-symmetric states}

The GHZ-symmetric states~\cite{ES2012}, denoted by $\rho^{\mathrm{S}}$, 
constitute the aforementioned peculiar family of symmetric states that facilitates 
the entire method described in this article. This is because the elements
of this family
contain much of the interesting physics of three-qubit states, 
but at the same time they are mathematically simple enough that 
many results can be calculated analytically.

\subsection{GHZ-symmetric three-qubit states}
The GHZ symmetry comprises the operations under which the well-known
GHZ state
\[
      \ket{\mathrm{GHZ}}\ = \oost\left( \ket{000}\ +\ \ket{111} \right)
\]
remains invariant, that is,
\begin{enumerate}
\item[{\em (i)}] qubit permutations, 
\item[{\em (ii)}] simultaneous three-qubit flips 
      (i.e., application of  $\sigma_x\otimes\sigma_x\otimes \sigma_x$), 
\item[{\em (iii)}] qubit rotations about the $z$ axis of the form
\[    U(\phi_1,\phi_2) = \rme^{\rmi \phi_1 \sigma_z}\otimes\rme^{\rmi \phi_2 \sigma_z}\otimes\rme^{-\rmi
      (\phi_1+\phi_2) \sigma_z}\ \ .
\]
\end{enumerate}
Here, $\sigma_x$ and $\sigma_z$ are Pauli matrices. The only pure states 
invariant under this symmetry are 
\begin{equation}
      \ket{\mathrm{GHZ}_{\pm}}\ \equiv\ \oost\left( \ket{000}\ \pm\ \ket{111} \right)
\end{equation}
while all other GHZ-symmetric states are mixed and, in the computational 
basis, are $8\times 8$ matrices of the shape
\begin{equation}   \rhoS\ \longleftrightarrow\ 
                                 \left(\begin{array}{ccccccc}
                                        a & & & & & & x\\
                                          & b \\
                                          & & .\\
                                          & & & . \\
                                          & & & & . \\
                                          & & & & &  b\\
                                        x & & & & & &  a
                                       \end{array}
                                 \right)
\label{eq:shape}
\end{equation}
with $a,b,x\in \mathbb{R}$. As $\tr\rhoS = 1$ there are effectively 
two real parameters that characterize the entire family.
A particularly nice (though not obvious) choice of parameters is
\begin{eqnarray}
  \label{eq:rhoparams}
 x(\rho^{\mathrm{S}}) &=&\! \frac{1}{2}\!
  \left[\bra{\mathrm{GHZ}_+}\rho^{\mathrm{S}}\ket{\mathrm{GHZ}_+}\! -\! 
        \bra{\mathrm{GHZ}_-}\rho^{\mathrm{S}}\ket{\mathrm{GHZ}_-}\right]\ \\
 y(\rho^{\mathrm{S}}) &=&\! \frac{1}{\sqrt{3}}
  \left[\bra{\mathrm{GHZ}_+}\rho^{\mathrm{S}}\ket{\mathrm{GHZ}_+}\ +\ \right.
\nonumber
\\ 
      && \ \ \ \ \ \ \ \ \ \ \ \ \ \ \ +\ \left.
        \bra{\mathrm{GHZ}_-}\rho^{\mathrm{S}}\ket{\mathrm{GHZ}_-} - \frac{1}{4}\right]
\end{eqnarray}
such that the geometry in the space of density matrices
induced by the Hilbert-Schmidt metric 
\[
     d^2_{\text{HS}}(A,B)
               \ \equiv
                                      \ \frac{1}{2}\tr(A-B)^{\dagger}(A-B)
\]
coincides with the 
Euclidean geometry of the $xy$ plane in Fig.~1.
For the parameters in Eq.~\eqref{eq:shape} we have $a=\frac{1}{8}+\frac{\sqrt{3}y}{2}$
and $b=\frac{1}{8}-\frac{y}{2\sqrt{3}}$.

As is well known there are two classes of three-qubit entangled states,
the GHZ class and the $W$ class~\cite{Duer2000,Acin2001}. In Ref.~\cite{ES2012}
it was shown that the GHZ/$W$ line, i.e., the border between these two classes can be calculated
exactly for GHZ-symmetric states (cf.~Fig.~1). The corresponding parametrized curve is
\begin{equation}
    x^W=\frac{v^5+8v^3}{8(4-v^2)}\ \ \ ,\ \ \
    y^W=\frac{\sqrt{3}}{4}\frac{4-v^2-v^4}{4-v^2}
\label{eq:b-W}
\end{equation}
with $-1\leqq v\leqq 1$. 
\begin{figure}
\centering
\includegraphics[width=.97\linewidth]{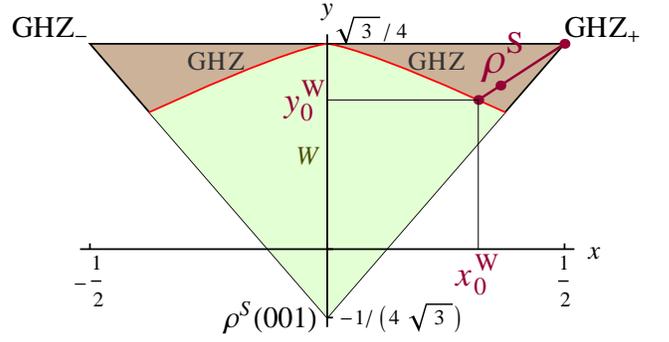}
    \caption{The family of GHZ-symmetric states $\rhoS$. 
             In the upper corners, there are the (pure)
             states GHZ$_{\pm}$ whereas the lower corner
             represents the separable mixture
    $\rhoS_{001}= \frac{1}{6} \sum_{jkl=001}^{110}\ \ket{jkl}\!\bra{jkl}$.
             The GHZ-type
             states ('GHZ') are separated from the states of at most
             $W$ type ('$W$') by the GHZ/$W$ line~\eqref{eq:b-W} (red
             solid line).
             Note the symmetry 
             $x \longleftrightarrow -x$ of the
             entanglement-related properties. 
\\
             For a GHZ-type state $\rhoS=\rhoS(x_0,y_0)$ we indicate the optimal
             decomposition for the three-tangle (cf.~Eq.~\eqref{eq:result3})
             consisting of the state GHZ$_+$ and the mixed $W$-type state
             $\rhoS(x_0^W,y_0^W)$.
            }
\end{figure}

\subsection{Exact three-tangle}

Coffman {\em et al.} discovered an entanglement measure for three qubits
that distinguishes between GHZ-type and $W$-type states, the residual
tangle~\cite{CKW2000}. As mentioned before, taking the square root of this quantity
has many advantages, and we refer to it as the {\em three-tangle}. 
For pure states it is defined as
\begin{eqnarray}
  \label{SIeq:three-tangle}
  \tau_3(\psi) &=& 2\sqrt{\left|d_1 - 2 d_2 + 4 d_3\right|}, 
  \nonumber
  \\
  d_1 &=& \psi_{000}^2\psi_{111}^2 + \psi_{001}^2\psi_{110}^2 + \psi_{010}^2\psi_{101}^2
  + \psi_{011}^2\psi_{100}^2                                
  \nonumber
  \\
  d_2 &=& \psi_{000}\psi_{001}\psi_{110}\psi_{111} + \psi_{000}\psi_{010}\psi_{101}\psi_{111} +
 \nonumber
 \\ &&
  + \psi_{000}\psi_{011}\psi_{100}\psi_{111}
  + \psi_{001}\psi_{010}\psi_{101}\psi_{110} + 
 \nonumber
 \\ &&
  + \psi_{001}\psi_{011}\psi_{100}\psi_{110} +
  \psi_{010}\psi_{011}\psi_{100}\psi_{101}
 \nonumber
 \\
  d_3 &=& \psi_{000}\psi_{110}\psi_{101}\psi_{011} + \psi_{100}\psi_{010}\psi_{001}\psi_{111}
\ \ .  
\end{eqnarray}
Here $\psi_{jkl}$ with $j,k,l\in\{0,1\}$ are the components of a pure three-qubit state
in the computational basis. It is easily checked that
\[
    \tau_3(\mathrm{GHZ})=1\ \ \mathrm{and} \ \ \ \ \tau_3(W)=0
\]
where $\ket{W}=\frac{1}{\sqrt{3}}(\ket{001}+\ket{010}+\ket{100})$.

The three-tangle for mixed states is more complicated, as it is defined as a 
convex roof~\cite{Uhlmann1998} 
\begin{equation}
      \tau_3(\rho)\ =\ \min_{\mathrm{\tiny all\ decomp.}} \sum\ p_j \ \tau_3(\psi_j) \ \ ,     
\label{eq:convex_roof}
\end{equation}
i.e., the minimum average three-tangle
taken over all possible pure-state decompositions $\{p_j,\psi_j\}$ for
$\rho=\sum_j p_j\ket{\psi_j}\!\bra{\psi_j}$. This is what makes the computation
of the three-tangle for mixed states difficult.
For GHZ-symmetric three-qubit states, however, the convex roof of the
three-tangle can be found exactly as~\cite{SE2012}
\begin{equation}
   \tau_3(x_0,y_0) =
  \begin{cases} \ \ \
    0 \ \ \ \ \ \ \ \ \ \ \mbox{for } x_0<x^W_0\ \mbox{and }y_0<y^W_0\\[2mm]
   \displaystyle
   \frac{x_0-x^W_0}{\frac{1}{2}-x_0^W}=
   \frac{y_0-y^W_0}{\frac{\sqrt{3}}{4}-y^W_0}  
                  \ \ \ \ \ \           \mbox{otherwise}
\ \ .
  \end{cases}
   \label{eq:result3}
\end{equation}
Here $x_0\geqq 0$
and $(x_0^W,y_0^W)$ are the coordinates of the intersection of
the GHZ/$W$ line with the direction that contains both GHZ$_+$ 
and $\rho^{\mathrm{S}}(x_0,y_0)$ (cf.~Fig.~1). 
The surface in Fig.~2 arises by
connecting each point of the GHZ/$W$ line $(x^W,y^W,\tau_3=0)$ 
with the closest of the points 
$(x_{\mathrm{GHZ}_{\pm}}=\pm\frac{1}{2},
  y_{\mathrm{GHZ}_{\pm}}=\frac{\sqrt{3}}{4},\tau_3=1)$.
That is, it interpolates linearly between the points of the
GHZ/$W$ line, and the maximally entangled states GHZ$_{\pm}$.
\begin{figure}
\centering
\includegraphics[width=.97\linewidth]{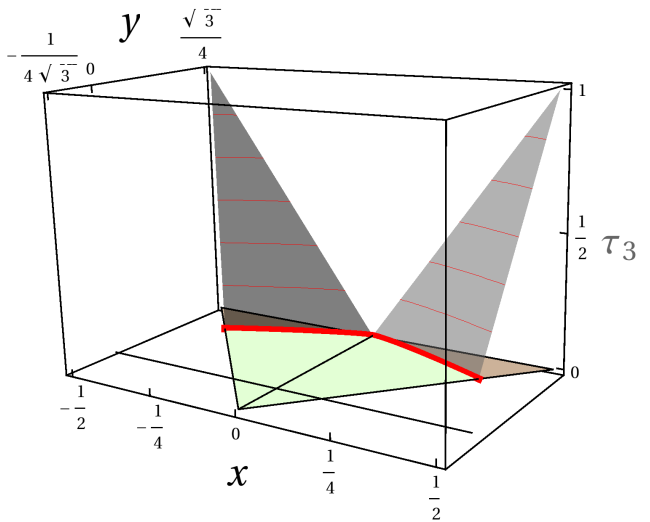}
    \caption{The exact three-tangle for the
             family of GHZ-symmetric states $\rhoS$, cf.~Eq.~\eqref{eq:result3}.
            }
\end{figure}

%

\subsection{An analytical approximation}
%

We note that the non-vanishing curvature of the GHZ/$W$ line Eq.~\eqref{eq:b-W}
is at the origin of the difficulty in writing down a more explicit formula 
than Eq.~\eqref{eq:result3} for $\tau_3(\rhoS)$. However, it is possible
to find analytical approximations.

In analogy with the discussion in Ref.~\cite{ES2012-ScR} we find a
plane that approximates the surface in Eq.~\eqref{eq:result3}.
The evident choice is a plane that contains the point 
$\left(x=\frac{1}{2},y=\frac{\sqrt{3}}{4},\tau_3=1\right)$
and that intersects the $xy$ plane
in a straight line tangential to the GHZ/$W$ line.
Each such tangent is an optimal GHZ witness~\cite{ES2013}.
 Particularly simple tangents are given by the witnesses
($\pm$ for $x\gtrless 0$)
\begin{equation}
    \mathcal{W}_{\pm} = \frac{3}{4}\id-\ket{\mathrm{GHZ}_{\pm}}\!\bra{\mathrm{GHZ}_{\pm}}
                                  -\frac{3}{7}\ket{\mathrm{GHZ}_{\mp}}\!\bra{\mathrm{GHZ}_{\mp}}
\label{eq:optwit}
\end{equation}
that describe via $\tr\left(\rhoS\mathcal{W}_{\pm}\right)=0$ the tangents 
touching the GHZ/$W$ line in the points $\left(x=\pm\frac{3}{8},y=\frac{\sqrt{3}}{6}\right)$,
respectively. The planes that contain one such tangent and the
corresponding three-tangle point for GHZ$_{\pm}$ are easily found 
(see Fig.~3) and give
\begin{equation}
 \tau_3^{\mathrm{approx}}(x,y)\ =\ \max{\left(0,\ \frac{4}{7}\left[-4+4|x|+5\sqrt{3}y\right]
                                        \right)}
\label{eq:tau3approx}
\end{equation}
as our analytical approximation to Eq.~\eqref{eq:result3} for the three-tangle of GHZ-symmetric
states. It is exact 
for all states that lie on the lower edges of the triangle. In principle, each 
optimal witness of the type \eqref{eq:optwit} from Ref.~\cite{ES2013} gives 
rise to an approximation analogous to Eq.~\eqref{eq:tau3approx}. They differ 
in the lines (in the $xy$ plane) for which they become exact.

\section{Symmetrizing an arbitrary state}
%
Up to this point our discussion has been 
restricted to states $\rho^{\mathrm{S}}$
with the symmetries {\em (i)--(iii)}.
A common way to extend our methods to arbitrary (non-symmetric) states $\rho$ is by a applying the projection
$\mathbf{P}: \rho \mapsto \rho^{\mathrm{S}}$ onto the symmetric states.
The operation
\begin{equation}
\label{eq:symmstate}
     \mathbf{P}(\rho)  \
      = \ \int\rmd U_{\mathrm{GHZ}}\, U_{\mathrm{GHZ}} 
                           \rho\, U_{\mathrm{GHZ}}^\dagger
\end{equation}
is often referred to as a twirling operation~\cite{Vollbrecht2002}. 
It averages over all symmetry elements {\em (i)--(iii)}. 

The effect of the projection $\mathbf{P}$ is easy to see. The matrix elements
of the image of $\rho$ are
\begin{eqnarray}
    \rhoS_{000,000}\ &=&\ \rhoS_{111,111} = \frac{1}{2} \left( \rho_{000,000}+\rho_{111,111}  \right)
\nonumber\\
    \rhoS_{jkl,jkl}\ &=&\ \frac{1}{6} \sum_{jkl=001}^{110}\ \rho_{jkl,jkl}
\nonumber\\
    \rhoS_{000,111}\  &=&\ \rhoS_{111,000}\ =\  
         \frac{1}{2}\left(\rho_{000,111}+\rho_{111,000}\right)
\nonumber
    \\ 
    \rhoS_{jkl,mnr}\ &=&\  0\ \ \ \ \mathrm{for \ all \ other \ matrix \ elements}\ \ .
\nonumber
\end{eqnarray}
From this the coordinates of the projection image in the $xy$ plane of Fig.~1
are readily obtained~\cite{ES2012}:
\begin{eqnarray}
    x(\rho)\ & =&\ \frac{1}{2} \left( \rho_{000,111}+\rho_{111,000}  \right)
\\
    y(\rho)\ & =&\ \frac{1}{\sqrt{3}}\left(\rho_{000,000}+\rho_{111,111}-\frac{1}{4}\right)
    \ \ .
\end{eqnarray}
It is worth noting that these relations, inserted into 
Eq~\eqref{eq:tau3approx} give a simple explicit formula 
to approximate the three-tangle of an arbitrary mixed state
\begin{align}
 \tau_3^{\mathrm{approx}}(\rho)  \ =\ &
 \max\left(
      0\ , \left[\pm \frac{8}{7}(\rho_{000,111}+\rho_{111,000})+
     \right.\right.
\nonumber \\ & \left.\left.
                       +  \frac{20}{7}(\rho_{000,000}+\rho_{111,111})
                      -3\right] \right)\ \ .
\label{eq:tau3approxRho}
\end{align}
Clearly, the performance of this formula is better the closer
state $\rho$ is to a GHZ-symmetric state. However, the local
bases will not always be arranged this way.  Therefore, the result
for $\tau_3^{\text{approx}}$ can be improved by applying
local unitaries to $\rho$.

We have already mentioned that symmetrizing a state usually results in
a loss of three-tangle. In order to see this consider the GHZ state
$\frac{1}{\sqrt{2}}(\ket{100}+\ket{011})$. Symmetrization maps it
to the lower corner of the triangle in Fig.~1,  i.e., 
to a completely separable
state. Naturally we would like to avoid such entanglement losses. This is why
we have to optimize the state before projecting it. While optimization cannot 
exclude entanglement loss, it may reduce it substantially. In the example,
the optimization is particularly simple. Applying $\sigma_x$ (a local 
operation that does not change entanglement) to the first qubit yields GHZ$_+$
which preserves its full entanglement under the projection $\mathbf{P}$.
\begin{figure}
\centering
\includegraphics[width=.97\linewidth]{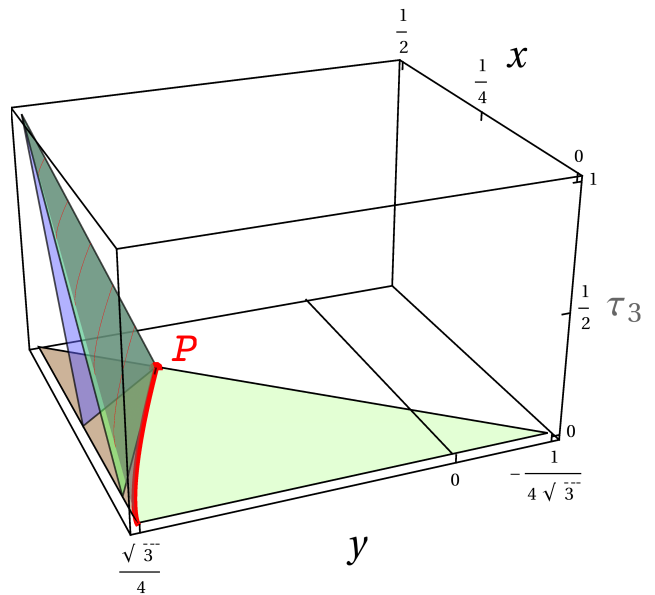}
    \caption{Simple quantitative witnesses for the three-tangle
             of GHZ-symmetric states $\rhoS$.
             In this figure, GHZ$_+$ is located in the left upper corner
             of the $xy$ plane. The point $P$ has the coordinates
             $(x=\frac{3}{8},y=\frac{\sqrt{3}}{6})$. 
             The upper-most tilted plane (light blue) represents the quantitative
             witness $\tau_3=-4\tr\left(\mathcal{W}\rhoS\right)$ where 
$
             \mathcal{W}=\frac{3}{4}\id_8-\ket{\text{GHZ}}\!\bra{\text{GHZ}}
$
             is the well-known projector-based GHZ witness~\cite{Acin2001}.
             An obviously better witness is $\mathcal{W}_+$, Eq.~\eqref{eq:optwit}.
             The corresponding plane (green), Eq.~\eqref{eq:tau3approx},
             contains the tangent
             to the GHZ/$W$ line in the point $P$ and is rather close
             to the exact solution Eq.~\eqref{eq:result3} (the dark green surface
             in the foreground). 
            }
\end{figure}

%

\section{Optimizing the state}
The example in the preceding section shows that projection onto the 
GHZ-symmetric states may serve to quantitatively assess the 
three-tangle of a state, and that optimization prior to projection
may enhance the reliability of the result. Clearly, symmetrization
can only {\em reduce}, and never {\em enhance} the entanglement of a
state~\cite{ES2012}. Therefore, if the projection image $\rhoS$
contains three-tangle, then the original state $\rho$
also does. That is why we also call this method a {\em 
quantitative witness}. The idea that witnessing entanglement 
can be improved by optimization was discussed by
various authors, e.g., in
Refs.~\cite{Bennett1996,Guehne2007,Eisert2007
}.

At first glance it looks difficult to find generally valid criteria
to improve an arbitrary state. In this section, we sketch why
the optimization steps {\em (1)} and {\em (2)} are appropriate for achieving
this goal.
%
\subsection{Normal form}

The normal form of a multipartite quantum state is a fundamental concept 
that was introduced by Verstraete {\em et al.}~\cite{Verstraete2003} and
discussed in detail also in Ref.~\cite{Leinaas2006}.
It applies to arbitrary (finite-dimensional) multi-qudit states. Here we focus
on $N$-qubit states only.
The essential merit of the normal form in the present context is that
among all states that are locally equivalent to the original state,
it is the one that maximizes a certain type of entanglement measures 
such as the three-tangle.
Let us elaborate on this point.

The defining property of the normal form $\rho^{\mathrm{NF}}$ 
is that {\em all} local density matrices are proportional to the identity
\[
    \left(\rho^{\text{NF}}\right)_{(j)}\ =\ 
         \tr_{1\ldots (j-1)(j+1)\ldots N} \rho^{\mathrm{NF}} \ \propto \ \id_2
    \ \ . 
\]
Therefore the normal form is unique only up to local unitaries.
Local equivalence to the original state $\rho$ means that 
$\rho^{\mathrm{NF}}$ can be obtained from $\rho$ by applying local operations 
\[
     \rho^{\mathrm{NF}}\ =\ \textstyle{(A_1\otimes\ldots\otimes A_N)}\rho
                            \textstyle{(A_1\otimes\ldots\otimes A_N)^{\dagger}} \ \ .
\]

Here the $A_j$ are {\em invertible} single-qubit operations that are not necessarily 
norm-preserving. For the normal form we additionally require $\det A_j =1$, or
more technically, $A_j \in \mathrm{SL}(2,\CC)$. 
Note that sometimes the normal form can be reached only 
  asymptotically. This is especially true for all pure states with
  vanishing three-tangle, where the normal form is simply zero.

How can we practically find the normal form? There is a simple iterative
procedure described in Ref.~\cite{Verstraete2003}. Let us denote the reduced local
density matrix of the $j$th qubit by $\rho_{(j)}$. One starts by transforming $\rho$
according to
\[
     \rho \to \frac{1}{\det \rho_{(1)}^{-1/2}} 
                            \left(\rho_{(1)}^{-1/2}\otimes \id_2\otimes\ldots\right)
                              \rho 
                       \left(\rho_{(1)}^{-1/2}\otimes \id_2\otimes\ldots\right)^{\dagger}   
\]
which brings $\rho_{(1)}$ into a form $\propto \id_2$. Next 
one applies the analogous
step to the second qubit. Note that, while making the second qubit $\propto \id_2$ 
this property on the first qubit is usually destroyed. 
Then one continues with the third qubit, and so on, to
the $N$th qubit, and then all over again. The convergence of this procedure 
was proved in Ref.~\cite{Verstraete2003}, and the result is the normal form. In 
each step of the iteration the trace is reduced (or  unchanged). As the three-tangle
remains unchanged under a transformation $A_j$ this means that, after renormalization
of the transformed state, the three-tangle has {\em increased}.
This is the reason why the normal form is a useful first optimization step 
for our purposes. If we normalize the normal form the resulting state
$\rho^{\mathrm{NF}}/\tr\rho^{\mathrm{NF}}$ has maximal
three-tangle in the local orbit of $\rho$~\cite{Verstraete2003}.

It is worth noticing that GHZ-symmetric
states---which play a central role in our discussion---are naturally given
in their normal form. 

\subsection{Criteria for unitary optimization}

As the normal form is unique only up to local unitaries another optimization
is required in order to find the appropriate local qubit bases.
That is, we transform the normalized state 
$\rho^{\mathrm{NF}}/\tr\rho^{\mathrm{NF}}$ by applying a unitary 
operation $V\in \mathrm{SU(2)}^{\otimes 3}$
\[
   \tilde{\rho}^{\mathrm{NF}}\ =\ V\ 
         \frac{\rho^{\mathrm{NF}}}{\tr\rho^{\mathrm{NF}}}\ V^{\dagger}\ \ .
\]
The obvious criterion is to maximize the three-tangle
\begin{equation*}
                         \tau_3\left(\mathbf{P}\left(
                                       \tilde{\rho}^{\mathrm{NF}}
                               \right)\right)
\ \longrightarrow\ \max\ \ ,
\end{equation*}
that is, 
\begin{equation}
   \tilde{\rho}^{\mathrm{NF}}\ =\ \argmax_{
                                      \scriptstyle V\in 
                                      \mathrm{SU(2)}^{\otimes 3}}\
                         \tau_3\left(\mathbf{P}\left(
                 V\ \frac{\rho^{\mathrm{NF}}}{\tr\rho^{\mathrm{NF}}}\ V^{\dagger}
                                        \right)\right)\ \ .
\label{eq:crit1}
\end{equation}
Clearly Eq.~\eqref{eq:crit1} gives the largest possible value of $\tau_3$.
However, due to the complicated structure of the function in
Eq.~\eqref{eq:result3} it appears hopeless to obtain analytical results.

Alternatively one may choose optimization criteria that do not necessarily give
the maximal value of $\tau_3$ but, depending on the state $\rho$, can 
possibly be treated analytically. Examples are the overlap 
(fidelity) with the GHZ state
\begin{equation}
   \tilde{\rho}^{\mathrm{NF}}_{\text{(f)}}\ =\ \argmax_{
                                      \scriptstyle V\in 
                                      \mathrm{SU(2)}^{\otimes 3}}\
        \bra{\mathrm{GHZ}}\ 
        V \frac{\rho^{\mathrm{NF}}}{\tr\rho^{\mathrm{NF}}}\ V^{\dagger}
        \ket{\mathrm{GHZ}} 
\end{equation}
or the Hilbert-Schmidt distance from the GHZ state
\begin{equation}
   \tilde{\rho}^{\mathrm{NF}}_{\text{(d)}}\ =\ \argmin_{
                                      \scriptstyle V\in 
                                      \mathrm{SU(2)}^{\otimes 3}}\
        d_{\text{HS}}\left(\pi_{\text{GHZ}}\ ,\
        V \frac{\rho^{\mathrm{NF}}}{\tr\rho^{\mathrm{NF}}}\ V^{\dagger}
                     \right)
\end{equation}
where $\pi_{\text{GHZ}}\equiv\ket{\text{GHZ}}\!\bra{\text{GHZ}}$.

\section{Experimental aspects}
In order to apply our method to an experimentally prepared state
its density matrix needs to be determined.
This is done using quantum state tomography~\cite{Paris2004}.
To this end we note that the density matrix can be
represented in a basis of local Pauli operators
\begin{equation}
      \rho\ =\ \frac18\sum_{jkl}\hat{\rho}_{jkl}\sigma_j\otimes\sigma_k\otimes
                                \sigma_l
\label{eq:tomog}
\end{equation}
where 
$\hat{\rho}_{jkl}=\tr\left(\rho\sigma_j\otimes\sigma_k\otimes\sigma_l\right)$.
Each term in Eq.~\eqref{eq:tomog} corresponds to a joint measurement
of local observables. To obtain
the optimized lower bound for the three-tangle (as described in the
preceding section) full tomography is required.
However, even 
in a small system like three qubits, full tomography requires 
the measurement of  63 observables (+ \tcb{possible} extra measurements for
normalization) and is therefore expensive.
There are ways to reduce the effort, 
as it is often the case that the experimentally prepared  state is not
completely unknown.

The first possibility is related to the observation that the 
analytical approximation 
Eq.~\eqref{eq:tau3approxRho} contains only the four entries
$\rho_{000,000}$, $\rho_{000,111}$, $\rho_{111,000}$, and $\rho_{111,111}$.
In Ref.~\cite{Guhne2007} the authors describe a general method to
determine the fidelity with the GHZ state. For three qubits
it requires four measurement settings. Those settings determine
the diagonal elements and the real part of $\rho_{000,111}$
separately, so that they are indeed sufficient to calculate the
projection of the state onto the GHZ symmetric states.
If the off-diagonal matrix element is not known to be real, it is
worthwhile determining the imaginary part as well and to use the
absolute value instead of the real part (this is equivalent to a
very restricted unitary optimization on the results). It turns out
that the additional setting $YYY$ (i.e., measurement of $\sigma_y$ for
all three qubits) is sufficient for that purpose.

Note that the best choice of local bases is such that the fidelity
of the GHZ state in those bases is maximized. In a sense, this corresponds
to carrying out the local unitary optimization for 
Eq.~\eqref{eq:tau3approxRho} directly in the experiment while it would
not be possible numerically, as not all elements of the density matrix
are known.

Another alternative to full tomography is permutationally
invariant tomography (PIT)~\cite{Toth2010}.
The experimental setup of PIT effectively means to apply the permutation
averaging part of the twirling operation (described in Section~II,
Eq.~\eqref{eq:symmstate}) directly in the
experiment, before the measurement. 
Therefore using PIT will in general result in a worse lower
bound. However, if the true state has approximate
permutation symmetry---which will often be the case in
experiments aiming at GHZ states---one may hope that
only little entanglement information is lost
in the projection $\rho\stackrel{\text{PIT}}{\longrightarrow}
\rho^{\text{(PI)}}$. 
For three qubits  10 settings are required~\cite{Toth2010}.
As the result of PIT is a valid density matrix
the optimization procedure may be applied to $\rho^{\text{(PI)}}$
as well to improve the estimate of $\tau_3$.

Also in the case of performing PIT it is possible to 
reduce the experimental effort by only
measuring the four GHZ matrix elements. 
If we content ourselves with the real part of the
measurement only, the three settings
$ZZZ$, $XXX$ and $XYY$ are sufficient. 
The real part of the off-diagonal element is
determined from 
$\langle XXX\rangle_{PIT} = \langle XXX\rangle$ and 
$\langle XYY\rangle_{PIT} =
\frac13\langle XYY+YXY+YYX\rangle$ as $\Re(\rho_{000,111}) = \langle XXX\rangle_{PIT} - 3\langle XYY\rangle_{PIT}$.
The imaginary part requires the additional settings
$XXY$ and $YYY$, so that 
$\Im(\rho_{000,111})=\langle YYY\rangle_{PIT} - 3\langle XXY\rangle_{PIT}$.

In summary, we see that there is a trade-off between the
experimental effort and the quality of the lower bound for $\tau_3$. 
Obviously the
best estimate is achieved by doing full tomography of the state and
then applying the procedure of Section~II to the result. 
In this case,
the best possible lower bound with our method is achieved. 

In the opposite extreme case, one measures the 
GHZ matrix elements $\rho_{000,000}$,
$\rho_{111,111}$, $\Re(\rho_{000,111})=\Re(\rho_{111,000})$
with an experimental PIT setup using
just three measurement settings in a specific local basis,
however, at the cost that
no further optimization of the bound is possible 
(but limited pre-measurement
optimization by appropriately choosing the local measurement bases).
Two additional measurement settings allow to determine
also $\Im(\rho_{000,111})$, which may already improve the bound.

Between those two extreme possibilities there is full PIT 
(requiring much fewer settings than full tomography, 
but maintaining the possibility to numerically optimizing 
the bound for $\tau_3$).
We note that---since the three-setting  and/or five-setting measurements
use the same setup
and a strict subset of the measurements for full PIT---they can be 
carried out 
first, and if this is not sufficient for a reasonable lower bound, one
can proceed by doing full PIT without wasting experimental effort.

%
With standard tomography, the minimal effort is four measurement
settings, which is slightly more than in a PIT setup. If the
imaginary part of the off-diagonal matrix is desired as well, only
one additional setting is required, so standard and PIT setups are
equally efficient for that case.

\section{Estimating the error of the lower bound}
As, in general, 
entanglement is lost in the projection 
$\mathbf{P}\left(\tilde{\rho}^{\text{NF}}\right)
\equiv \tilde{\rho}^{\text{S}}$
it is desirable to get an idea by how much 
$\tau_3\left(\tilde{\rho}^{\mathrm{S}} \right)$
underestimates the three-tangle of 
$\tilde{\rho}^{\text{NF}}$.
Recall that we know the value $\tau_3(\tilde{\rho}^{\text{S}})$
{\em exactly} and, moreover, 
$\tilde{\rho}^{\text{S}} \leqq \tilde{\rho}^{\text{NF}}$. 
However, we do not know how much larger 
$\tau_3\left(\tilde{\rho}^{\text{NF}}\right)$
actually is. Consequently we need an {\it upper}
bound to $\tau_3\left(\tilde{\rho}^{\text{NF}}\right)$
to estimate the error.

In principle, any available decomposition of $\tilde{\rho}^{\text{NF}}$
provides an upper bound to $\tau_3\left(\tilde{\rho}^{\text{NF}}\right)$.
The simplest upper bound is 
$\tau_3\left(\tilde{\rho}^{\text{NF}}\right)\leqq 
\bra{\text{GHZ}} \tilde{\rho}_{\text{(f)}}^{\text{NF}}\ket{\text{GHZ}}$.
A good upper bound can be found numerically using elaborate 
methods as described in Refs.~\cite{Loss2008,Cao2010,Zhu2012,Love2013}.
Alternatively, one can refine the simple estimate
by applying  a geometrical method that 
represents a variant of the ideas in Ref.~\cite{Love2013}.
It uses the convexity of both the state space and of the three-tangle
and is most easily explained by considering Fig.~4.

\begin{figure}[h]
\centering
\includegraphics[width=1.\linewidth]{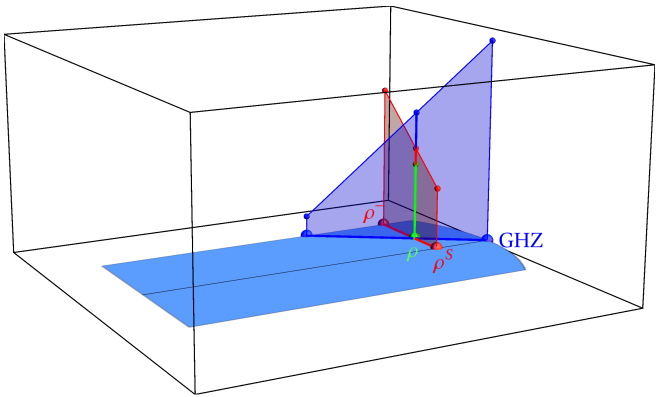}
    \caption{Illustration for the error estimate. The base area (light-blue)
             represents the set of all three-qubit states. The (black solid) line
             in it corresponds to the GHZ-symmetric states and has
             the state GHZ$_+$ at its right end point. A density matrix $\rho$
             (light green) is projected by the twirling
             operation Eq.~\eqref{eq:symmstate}
             onto $\rhoS$ (red) in the GHZ-symmetric states,
             thereby reducing the three-tangle (which is displayed in
             the vertical direction). The state $\rho_-$ is found by
             extending the straight line from $\rhoS$ through $\rho$
             towards the border of the state space. An upper bound 
             $\tau_3^-\geqq\tau_3(\rho_-)$ 
             yields an upper bound for $\tau_3(\rho)$
             due to the convexity of the three-tangle. Alternatively,
             one may use any other state instead of $\rhoS$ with known three-tangle
             (for example, the GHZ state, as indicated by
             the other vertical plane.
            }
\end{figure}

In the figure, the GHZ-symmetric states are represented by a
line. They form a subset of the complete state space. The
projection of the optimized state $\tilde{\rho}^{\text{NF}}$
onto the GHZ-symmetric states is then just an orthogonal
projection onto that line.  We may assume
$\tilde{\rho}^{\text{NF}}\neq\tilde{\rho}^{\text{S}}$,
otherwise we do not need an error estimate.

Consider now the straight line connecting $\tilde{\rho}^{\text{S}}$
and $\tilde{\rho}^{\text{NF}}$. We extend this line until it 
reaches the border of the space of density matrices in the state $\rho_-$.
This means $\rho_-$ is an affine combination of
$\tilde{\rho}^{\text{S}}$ and $\tilde{\rho}^{\text{NF}}$
\[
      \rho_-\ =\ \frac{1}{\lambda}
                         \tilde{\rho}^{\text{NF}}\ -\
                 \frac{1-\lambda}{\lambda}
                         \tilde{\rho}^{\text{S}}
\]
with some real $\lambda\in (0,1]$ that is 
defined by the condition that the smallest eigenvalue of
$\rho_-$ be zero.

Given $\rho_-$ we can determine an upper bound 
$\tau_3^-\geqq\tau_3(\rho_-)$ and obtain 
\[ \tau_3\left(\tilde{\rho}^{\text{NF}}\right)\leqq 
   \lambda\ \tau_3^-\ +\ (1-\lambda)\
                   \tau_3\left(\tilde{\rho}^{\mathrm{S}} \right)
\ \ .
\]
Evidently this method to estimate 
$\tau_3\left(\tilde{\rho}^{\text{NF}}\right) $
can be applied analogously with any other state 
$\rho$ (in place of $\tilde{\rho}^{\text{S}}$) for which
the exact three-tangle is known, for example, the states GHZ$_{\pm}$.

\section{Examples}
In this section we illustrate and analyze the performance of our method
by applying it to states whose three-tangle is known exactly, or with
high numerical precision.
Among other details, this reveals that our method 
gives exact results not only for GHZ-symmetric states.

\subsection{Mixtures of a GHZ state with a product state}
Consider mixtures of the GHZ state and an orthogonal
product state
\begin{equation}
     \rho_{[1]}(p)\ =\ p\ \ket{\mathrm{GHZ}}\!\bra{\mathrm{GHZ}}
                \ +\ (1-p)\ \ket{\mathrm{001}}\!\bra{\mathrm{001}}
\end{equation}
that we mentioned briefly in the Supplement of Ref.~\cite{ES2012-ScR}.
The exact solution $\tau_3(\rho_{[1]}(p))=p$ can be derived 
following Refs.~\cite{EOSU2008,Viehmann2012}.
Here we give the analytical solution by means of the method 
discussed in the present article.

By applying the local operation
\[
       A_{[1]}\ =\
       \left(\begin{array}{cc}
                \alpha & 0\\
                   0   & 1/\alpha
             \end{array}
       \right)\
\otimes \ \id_2\ \otimes
       \left(\begin{array}{cc}
                1/\alpha & 0\\
                   0   & \alpha
             \end{array}
       \right)\
\]
to $\rho_{[1]}$ we obtain
\[
   A_{[1]} \rho_{[1]}A_{[1]}^{\dagger}\ =\ 
      p\ \ket{\mathrm{GHZ}}\!\bra{\mathrm{GHZ}}
                \ +\ (1-p)\ \alpha^4\ \ket{001}\!\bra{001}
\]
and letting $\alpha\to 0$ the normal form
\[
      \rho_{[1]}^{\text{NF}}(p)\ =\ 
      p\ \ket{\mathrm{GHZ}}\!\bra{\mathrm{GHZ}}\ \ .
\]
No further unitary optimization is required and the result is
\begin{equation}
   \tau_3(\rho_{[1]}(p))\ =\ \tr \rho_{[1]}^{\text{NF}}(p)\ 
                          \tau_3(\text{GHZ})\ =\ p\ \ ,
\end{equation}
showing that there are also cases of states for which the exact three-tangle
is obtained although they are not GHZ symmetric. This is not a 
coincidence. The states $\rho_{[1]}$ 
can be obtained from a two-qubit state simply by
doubling the first qubit in each basis state 
(that is, $\ket{jk}\!\bra{lm}\to \ket{jjk}\!\bra{llm}$).
Multipartite entanglement in such states
was analyzed in Ref.~\cite{EOS2009}. Consequently, these states belong to
a subspace that can be described by two-qubit physics. However, as
the method is exact for arbitrary two-qubit states~\cite{ES2012-ScR},
it must be exact also for $\rho_{[1]}$.

\subsection{Mixtures of a GHZ state and a $W$ state}
Our second example is
\begin{equation}
     \rho_{[2]}(p)\ =\ p\ \ket{\mathrm{GHZ}}\!\bra{\mathrm{GHZ}}
                    \ +\ (1-p)\ \ket{W}\!\bra{W}
\end{equation}
for which the exact solution was found in Ref.~\cite{LOSU2006}
(for the residual tangle) and in Ref.~\cite{Viehmann2012} (for
the three-tangle). The states contain three-tangle for 
$p\geqq p_0=\frac{ 2^{1/3}}{2^{1/3}+3/4}\approx 0.627$ so that
\[
     \tau_3(\rho_{[2]}(p))\ =\ \max\left(\ 0\ , \ \frac{p-p_0}{1-p_0}
                              \  \right)
\ \ .
\]
First we note that $\rho_{[2]}$ is not in normal form, however,
for $p\geqq p_0$ it is rather close to it. The local transformation to get the
normal form is diagonal and can, in principle, be obtained
analytically. To render the discussion more transparent we
will use the approximation
\[
       \rho_{[2]}^{\mathrm{NF}}(p)\ \approx\ \rho_{[2]}(p)\ \ .
\]
Then again, no unitary optimization is necessary and the
symmetrized state is approximately
\[
 \begin{aligned}
       \tilde{\rho}_{[2]}^{\mathrm{S}}(p)  \approx &\ 
             p\ \pi_{\text{GHZ}}\ +\ 
\\
        & + \frac{1-p}{6}\left(\pi_{001}
          +\pi_{010}+\pi_{100}+\pi_{011}+\pi_{101}+\pi_{110}\right)
 \end{aligned}
\]
where $\pi_{klm}\equiv\ket{klm}\!\bra{klm}$. The states 
$\tilde{\rho}^{\text{S}}_{[2]}(p)$ are located at the lower right border
of the triangle in Fig.~1. They have non-vanishing three-tangle for 
$p>3/4$ and, hence, 
\[
     \tau_3(\tilde{\rho}_{[2]}(p))\ \approx
     \ \max\left(\ 0\ , \ 4p-3 \
           \right)
\ \ .
\]
We clearly observe the loss of tripartite entanglement due to the
projection, in the worst case (for $p=3/4$) it amounts to
$\frac{3/4-p_0}{1-p_0}\approx 0.33$. Note also that the
error estimate of Section~VI is not helpful in a case like
this: Since $\rho_{[2]}$ is of rank two it is located at the
border of the state space for most directions. On the other
hand, a considerable part of the large errors occurs for states
$\rho_{[2]}$ that have non-zero three-tangle while for the
estimate $\tau_3(\tilde{\rho}^{\text{S}}_{[2]})=0$. In such cases,
a numerical method like that of Ref.~\cite{Love2013}
is highly useful as it is capable of certifying (numerically) vanishing
three-tangle.

\subsection{A nontrivial $W$ state}

The last example we consider in this section is the state
\begin{equation}
 \rho_{[3]}\ =\ \frac{1}{8}
                \left(
                       \begin{array}{cccccccc}
                         1 & 0 & 0 & 0 & 0 & 0 & 0 & 0 \\
                         0 & 1 & 1 & 1 & 1 & 1 & 1 & 0 \\
                         0 & 1 & 1 & 1 & 1 & 1 & 1 & 0 \\
                         0 & 1 & 1 & 1 & 1 & 1 & 1 & 0 \\
                         0 & 1 & 1 & 1 & 1 & 1 & 1 & 0 \\
                         0 & 1 & 1 & 1 & 1 & 1 & 1 & 0 \\
                         0 & 1 & 1 & 1 & 1 & 1 & 1 & 0 \\
                         0 & 0 & 0 & 0 & 0 & 0 & 0 & 1 
                       \end{array}
                \right)
\ \ .
\end{equation}
It is a rank-3 state, a mixture of 
the GHZ-type state $\ket{\varphi}=\frac{1}{\sqrt{6}}(\ket{001}+\ket{010}+
\ldots +\ket{110})$ and two product states $\pi_{000}$ and $\pi_{111}$.
It is difficult to decide at first glance which class $\rho_{[3]}$ belongs
to.

Application of the procedure described in Section~II yields,
after projection to the GHZ-symmetric states, a point in
the $W$ region. That is, the state is at least of $W$ type: The
state $\tilde{\rho}_{[3]}^{\text{NF}}$ before the projection
is locally equivalent to $\rho_{[3]}$ and the image of the projection
$\tilde{\rho}_{[3]}^{\text{S}}$ has at most the class of the original
state.  However, we cannot be sure that its three-tangle indeed
equals zero.

We may try to find an explicit decomposition of $\rho_{[3]}$
that has vanishing three-tangle.  It turns out that 
a decomposition into three pure states of the form 
\[
      \ket{\psi_j}\ =\ c_{j1}\ket{000}+c_{j2}\ket{\varphi}+c_{j3}\ket{111}
\]
is sufficient.
A numerical minimization indeed gives a decomposition with
numerically vanishing three-tangle:
\begin{eqnarray}
  \label{eq:decomp}
  \rho_{[3]}\ & \approx\ & 0.32809\ket{\psi_1}\!\bra{\psi_1}\ +
\nonumber\\
              &&  0.52694\ket{\psi_2}\!\bra{\psi_2}\
               +\ 0.14498\ket{\psi_3}\!\bra{\psi_3}
\end{eqnarray}
with
\begin{eqnarray*}
  \ket{\psi_1} &\approx& 0.45488\ket{000} - 0.86186\ket{\varphi} - 0.22423\ket{111}\\
  \ket{\psi_2} &\approx& 0.28991\ket{000} + 0.95673\ket{\varphi} - 0.02476\ket{111}\\
  \ket{\psi_3} &\approx& 0.29741\ket{000} - 0.40662\ket{\varphi} + 0.86383\ket{111}
\end{eqnarray*}
(For a precision better than $\tau_3<10^{-20}$ many more digits 
of the coefficients need to be taken into account.)

Consequently, our method works well also for states whose optimal 
decompositions are more intricate. A possible reason for the reliability
in the case of $\rho_{[3]}$ is the presence of both
permutation and spin flip invariance in the state.
%
%
%

\section{Summary} 
We have described the practical implementation of a method that yields
a lower bound---in principle, analytical---to the three-tangle of an
arbitrary three-qubit mixed state. While the theoretical grounds of this
method were investigated in Refs.~\cite{ES2012,SE2012,ES2012-ScR}
we have focused here on various practical aspects. Apart from the ingredients
of the method we have provided a more detailed discussion of the 
approximation to the three-tangle by projector-based entanglement witnesses
(and the corresponding quantitative witnesses), Section III.C, 
as well as an analytical lower bound for the three-tangle of 
an arbitrary three-qubit density matrix, Eq.~\eqref{eq:tau3approx}.
Moreover, we have considered the tradeoff between the quality of
the lower bound and the experimental effort to determine
elements of the density matrix.

As the procedure explained here gives only a lower bound to the three-tangle
one would like to know about the possible error. 
To this end, we have described  an error estimate
based on simple geometrical considerations. Finally, we have studied
the performance of the method by comparing its results to exactly
solvable cases for the three-tangle. From this we may conclude 
that the method often works well, however, in particular
for small three-tangle in the original state the bound may
substatially underestimate the entanglement. Therefore, it is
useful to combine it with a numerical approach (such as in Ref.~\cite{Love2013})
that can numerically certify vanishing three-tangle.
%
%
%
%

\section*{Acknowledgements}
This work was funded by the German Research Foundation within 
SPP 1386 (C.E.), by Basque Government grant IT-472-10 
and MINECO grant FIS2012-36673-C03-01 (J.S.).
The authors thank  P.J.\ Love for interesting discussions, 
C.\ Schwemmer and G.\ T\'oth for helpful comments, and
J.\ Fabian and K.\ Richter for their support.
%
%

%
%
%

\section*{references}

\end{document}